\documentclass[twocolumn,showpacs,prl,amsmath,amssymb,notitlepage,superscriptaddress]{revtex4-1}

\usepackage{graphicx}
\usepackage{color}
\usepackage{amsmath}

\begin{document}



\title{Abrupt phase transition of epidemic spreading in simplicial complexes}

\author{Joan T. Matamalas}
 \affiliation{%
 Departament d'Enginyeria Inform\`{a}tica i Matem\`{a}tiques, Universitat Rovira i Virgili, 43007 Tarragona, Spain.\\
}%
\author{Sergio G\'omez}
 \affiliation{%
 Departament d'Enginyeria Inform\`{a}tica i Matem\`{a}tiques, Universitat Rovira i Virgili, 43007 Tarragona, Spain.\\
}%
\author{Alex Arenas}
 \affiliation{%
 Departament d'Enginyeria Inform\`{a}tica i Matem\`{a}tiques, Universitat Rovira i Virgili, 43007 Tarragona, Spain.\\
}%


\date{\today}

\begin{abstract}
Recent studies on network geometry, a way of describing network structures as geometrical objects, are revolutionizing our way to understand dynamical processes on networked systems. Here, we cope with the problem of epidemic spreading, using the Susceptible-Infected-Susceptible (SIS) model, in simplicial complexes. In particular, we analyze the dynamics of SIS in complex networks characterized by pairwise interactions (links), and three-body interactions (filled triangles, also known as 2-simplices). This higher-order description of the epidemic spreading is analytically formulated using a microscopic Markov chain approximation. The analysis of the fixed point solutions of the model, reveal an interesting phase transition that becomes abrupt with the infectivity parameter of the 2-simplices.
Our results pave the way to advance in our physical understanding of epidemic spreading in real scenarios where diseases are transmitted among groups as well as among pairs, and to better understand the behavior of dynamical processes in simplicial complexes.
\end{abstract}


\maketitle


The collective behavior of dynamical systems on networks, has been a major subject of research in the physics community during the last decades~\cite{Strogatz2003,Arenas2008PhysRep,barrat2008,Newman2010,Pastor2015revepi}. In particular, our understanding of both natural and man-made systems has significantly improved by studying how network structures and dynamical processes combined shape the overall systems' behavior. Recently, the network science community has turned its attention to network geometry~\cite{Horak2009,wu2015,benson2018,Salnikov2019EJP} to better represent the kinds of interactions that one can find beyond typical pairwise interactions.

These higher--order interactions are encoded in geometrical structures that describe the different kinds of simplex structure present in the network: a filled clique of $m+1$ nodes is known as an $m$-simplex, and together a set of 1-simplexes (links), 2-simplexes (filled triangles), etc., comprise the simplicial complex. While simplicial complexes have been proven to be very useful for the analysis and computation in high dimensional data sets, e.g., using persistent homologies~\cite{Petri2014Interface,Giusti2016JCN,Ashwin2016PhysD,Reimann2017,Otter2017EPJDS}, little is understood about their role in shaping dynamical processes, save for a handful of examples~\cite{Gin2018,Gin2019,Iacopini2019NatComms,skardal2019}.

A more accurate description of dynamical processes on complex systems necessarily requires a new paradigm where the network structure representation helps to include higher-order interactions~\cite{Lambiotte2019}. Simplicial geometry of complex networks is a natural way to extend many-body interactions in complex systems. The standard approach so far consists in understanding the coexistence of two-body (link) interactions and three-body interactions (filled triangles). Note that this approach is different from considering pairwise interactions among three elements of a triangle, it refers to the interaction of the three elements, in the filled triangle, at unison.

Here we present a probabilistic formalization of the higher--order interactions of an epidemic process, represented by the well-known Susceptible-Infected-Susceptible (SIS) model~\cite{heat2000} in one and two-simplices, revealing that the consideration of higher-order structure (filled triangles) can change the character of the phase transition of the epidemic spreading to the endemic state. Specifically, we find that for a significant region of the parameters space the continuous phase transition that has been well-characterized in networks so far~\cite{Pastor2015revepi} becomes abruptly discontinuous~\cite{Rev2019}. These results are important for physicists working on network science, independently on the particular dynamical process we have chosen, to explain why this physical phenomena could arise in complex systems. Similar results have been reported for social contagion dynamics~\cite{Iacopini2019NatComms} using a mean-field approach.


For the mathematical formalization of the dynamical process, we use simplicial complexes extensions of the \textit{Microscopic Markov Chain Aproach} (MMCA)~\cite{gomez10, gomez11,granell2013,granell2014,david181,david182}, and of the \textit{Epidemic Link Equations} (ELE)~\cite{matamalas18}, that compute the probabilities of 1~and 2-simplexes to transmit the epidemics. This formalism allows us to get physical insight into the phase transition and its consequences at the level of our understanding of plausible epidemic scenarios.



%
Let us start by considering the dynamics of the SIS model in networks. We consider a network of $N$ nodes and $L$ links, where the nodes can have two states, susceptible (S) or infected (I). The classical interaction so far considers that the infection propagates, pairwise, from certain infected individuals to their neighbors with a probability $\beta$, and infected nodes eventually recover with probability $\mu$. In the simplicial complex scenario, we will consider also triangular interactions, i.e., every node also interacts within the 2-simplexes with the two neighbors at unison, with an infection probability $\beta^\triangle$ when the other two members are infected. We can define a system of $N$ discrete-time MMCA equations that capture the evolution over time of the probability $p_i$ of node~$i$ being in an infected state as:
\begin{equation}
  p_i(t + 1) = (1 - p_i(t))(1 - q_i(t)q_i^\triangle(t)) + p_i(t)(1 - \mu)\,,
  \label{eq:mmca_eq}
\end{equation}
where $q_i(t)$ defines the probability that node~$i$ is not infected by any pairwise interaction with its neighbors,
\begin{equation}
  \label{eq:q_i_mmca}
  q_i(t) = \prod_{j\in\Gamma_i}\left( 1 - \beta p_i(t) \right)\,,
\end{equation}
and $q_i^\triangle(t)$ is the probability that node~$i$ is not infected by any of its interactions at the 2-simplicial level,
\begin{equation}
  \label{eq:q_i_tri_mmca}
  q_i^\triangle(t) = \prod_{j, \ell \in \triangle_i}\left( 1 - \beta^\triangle p_j(t)p_\ell(t) \right)\,.
\end{equation}
Here, $\Gamma_i$ and $\triangle_i$ represent the sets of 1~and 2-simplexes containing node~$i$, respectively.



The system of equations Eq.~(\ref{eq:mmca_eq}) updates the probability of a node~$i$ being infected as the probability of being susceptible at time~$t$ and becoming infected by some neighbor, or some group of neighbors in a triangle (first term on the r.h.s.\ of the equation), or the probability that node~$i$ was already infected at time~$t$ and it does not recover. This system of equations is a contraction map $T_{\beta,\beta^{\triangle},\mu}: \vec{p}(t) \rightarrow \vec{p}(t+1)$ for every value of the parameters, and then the existence of fixed points is guaranteed by the Banach fixed point theorem~\cite{adam2013}. We can solve the system by iteration.
A na\"{\i}ve approximation of Eq.~(\ref{eq:mmca_eq}) is the homogeneous assumption in which all nodes have the same degree ($k$), belong to the same number of 2-simplexes ($k^{\triangle}$), and have the same probability of being infected ($p_i(t)=p(t))$. Expanding up to second order in $p$, and developing the equation at the stationary state ($p(t)=p$), it reveals the structure responsible for the abrupt transitions we could foresee:
\begin{equation}
  (1-p) \left[ \beta k p  + \left(\beta^{\triangle} k^{\triangle} - \beta^2 k (k-1)/2\right) p^2 \right] - \mu p = 0\,.
\label{eq:mfield}
\end{equation}

\begin{figure}[tb]
  \centering
  \includegraphics[width=\columnwidth]{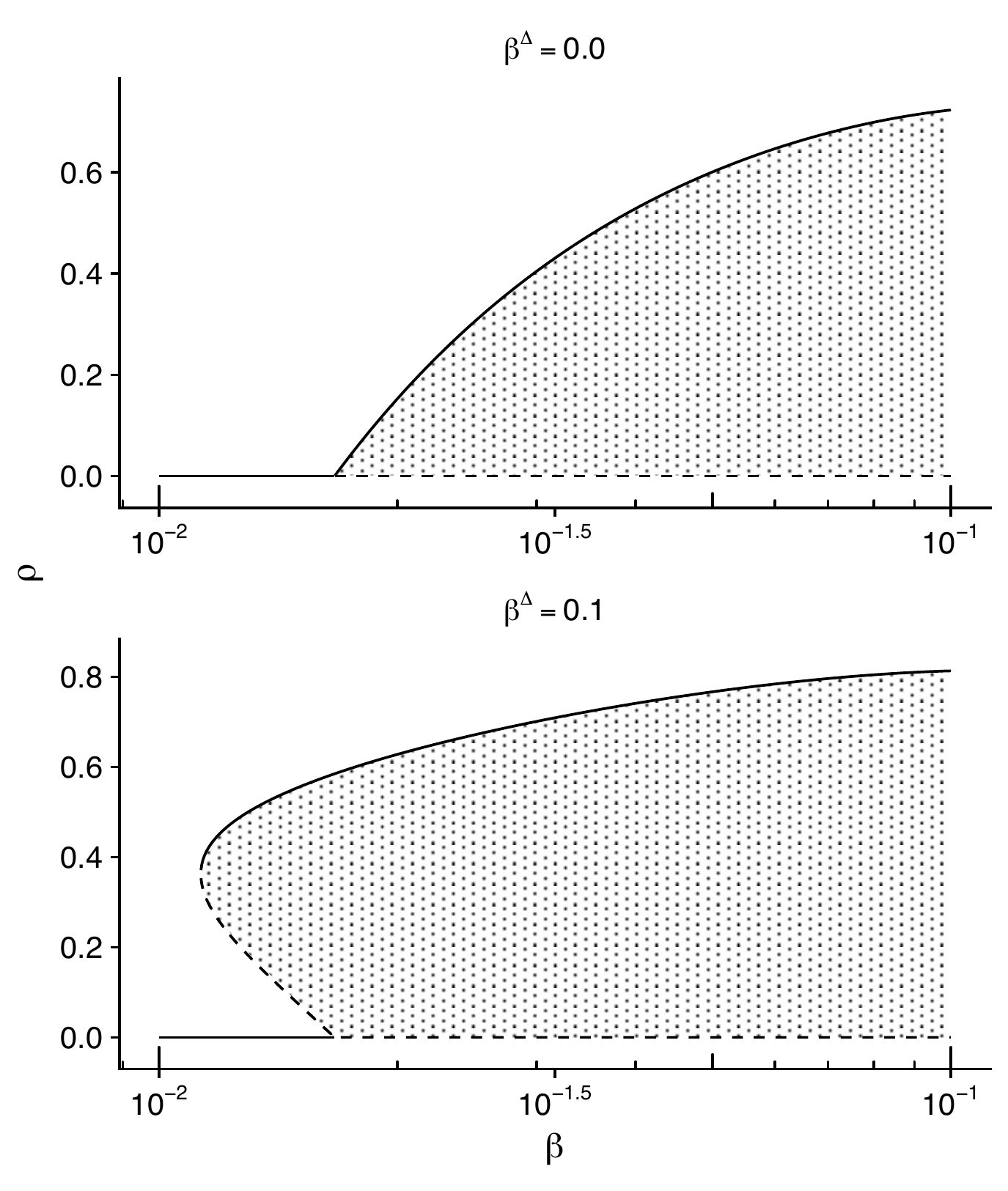}
  \caption{Phase diagram of the mean-field approximation detailed in Eq.~\eqref{eq:mfield}. We consider a network with average degree $\langle k \rangle = 12$, an average number of triangles that each node is part of $\langle k^{\triangle} \rangle = 5$, a recovery probability $\mu = 0.2$, and two different infection probabilities through ternary interactions, $\beta^{\triangle} = 0.0$ (top) and $\beta^{\triangle} = 0.1$ (bottom). Solid lines represent stable solutions and dashed lines depict unstable solutions to Eq.~\eqref{eq:mfield}. The dotted area represents the region of parameters that enables a convergence to an endemic stable state from below.}
  \label{fig:mf_stability}
\end{figure}

This 3rd-order algebraic equation has a trivial solution, $p=0$, and two more roots, that depending on the parameters,
provides up to two additional physical solutions with $p\in(0,1]$. For the parameters with three physical solutions, the stability analysis shows that the middle one is unstable, thus being responsible of the abrupt transitions, see Figure~\ref{fig:mf_stability}.
This result is equivalent to that obtained in~\cite{Iacopini2019NatComms}.

Note, however, that Eq.~(\ref{eq:mmca_eq}) and in turn Eq.~(\ref{eq:mfield}) carries the implicit assumption that the probability of being infected by one neighbor is independent on the probability of being infected by any other neighbor. This assumption is a mean-field approximation, whose validity is severely compromised in the current scenario, where we account for the infectivity in triads, and hence neighbors are unlikely to be independent.

To palliate the previous limitation, we can define the simplicial epidemic model at a level of links using a system of $3L$ equations. The size of this system is very large compared to the previous MMCA system of $N$ equations at the level of nodes; however, we can simplify the model to a system of $L + N$ equations if we consider the constrains imposed by the probabilistic model, i.e., the node marginal probabilities have to be the same regardless the link that we consider to compute them. These restrictions read as follows: for a link connecting nodes~$i$ and~$j$, the probability of a node~$i$ to be in the susceptible state is $P_i^S = P_{ij}^{SS} + P_{ij}^{SI}$, where $P_{ij}^{SS}$ is the joint probability of node~$i$ being susceptible and node~$j$ being susceptible, and $P_{ij}^{SI}$ is the joint probability of node~$i$ being susceptible and node~$j$ being infected. Equivalently $P_i^I = P_{ij}^{II} + P_{ij}^{IS}$. Wrapping up these restrictions we can write the Epidemic Link Equations, ELE~\cite{matamalas18}, for the simplicial model, for every node~$i$, as
%
\begin{equation}
  P_i^I(t + 1) = (1 - P_i^I(t))(1 - q_i(t) q_i^\triangle(t)) + P_i^I(t) (1 - \mu)\,,
  \label{eq:nodesELE}
\end{equation}
where $q_i(t)$ defines the probability that node~$i$ is not infected by any pairwise interaction with its neighbors,
\begin{equation}
  q_i(t) = \prod_{j\in\Gamma_i}\left(1 - \beta\frac{P_{ij}^{SI}(t)}{P_i^S(t)}\right)\,,
\end{equation}
and $q_i^\triangle(t)$ is the probability that node~$i$ is not infected by any of the interactions at the 2-simplicial level,
\begin{equation}
  q_i^\triangle =\prod_{j,r \in \triangle_i}\left(1 - \beta^\triangle \frac{P_{ijr}^{SII}(t)}{P_i^S(t)}\right)\,.
\end{equation}
Note that, to write down the previous equations, we have made use of Bayes' theorem, substituing the conditional probabilities $P_{ji}^{I|S}$ of node~$j$ to be infected knowing that node~$i$ is susceptible, by the joint probability $P_{ij}^{SI}/P_i^S$; equivalently, for the 2-simplicial terms, $P^{II|S}_{jri} = P_{ijr}^{SII}/P_i^S$.

\begin{figure}[tb]
  \centering
  \includegraphics[width=0.8\columnwidth]{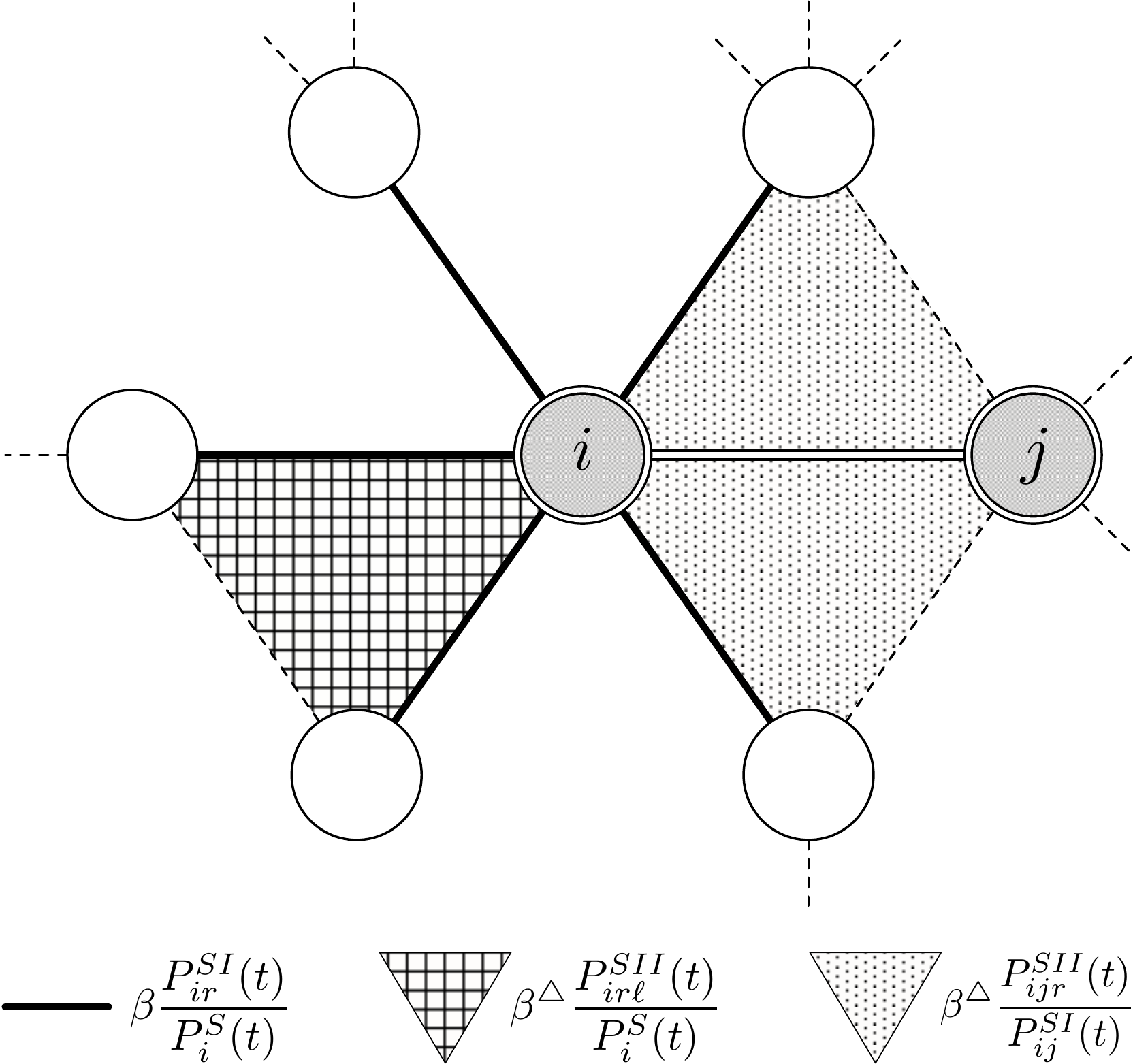}
  \caption{Schematic representation of the contribution of different interactions to the joint probability between the states of node~$i$ and node~$j$, through node~$i$. Black solid lines account for direct interactions between node~$i$ and its neighbors. The checked pattern area represents a 2-simplex interaction with node~$i$ where node~$j$ is not participating, and the dotted pattern areas represent 2-simplex interactions with node~$i$ where node~$j$ is participating. All these contributions participate in Eqs.~(\ref{eq:qij}) to~(\ref{eq:uij}), respectively.}
  \label{fig:schema_bw}
\end{figure}

The system still requires of $L$~equations, one for every link, that account for the probability of having a link connecting two nodes in the infected state $II$, transitioning from the four possible states $SS$, $SI$, $IS$, $II$. It reads
\begin{eqnarray}
  P_{ij}^{II} (t&+&1) =  P_{ij}^{SS}(t)\left(1 - q_{ij}(t)q_{ij}^\triangle(t)\right)\left(1 - q_{ji}(t)q_{ji}^\triangle(t)\right) + \nonumber \\
  & + & P_{ij}^{SI}(t)\left(1 - (1 - \beta)q_{ij}(t)u^\triangle_{ij}(t)q_{ij}^\triangle(t)\right)(1 - \mu) + \nonumber \\
  & + & P_{ij}^{IS}(t)\left(1 - (1 - \beta)q_{ji}(t)u^\triangle_{ji}(t)q_{ji}^\triangle(t)\right)(1 - \mu) + \nonumber \\
  & + & P_{ij}^{II}(t)\left(1 - \mu\right)^2\,,
  \label{eq:linksELE}
\end{eqnarray}
where we have used: the probability of node~$i$ not being infected by any neighbor different from~$j$ through a link,
\begin{equation}
  q_{ij}(t) = \prod_{\substack{{r\in\Gamma_i} \\ r \neq j}}
              \left(1 - \beta\frac{P^{SI}_{ir}(t)}{P^S_i(t)}\right)\,;
  \label{eq:qij}
\end{equation}
the probability of node~$i$ not being infected by any 2-simplex not containing node~$j$,
\begin{equation}
  q^\triangle_{ij}(t) = \prod_{\substack{r,\ell \in \triangle_i \\ r \neq j \\ \ell \neq j}}
                        \left(1 - \beta^\triangle \frac{P^{SII}_{i r\ell}(t)}{P_i^S(t)}\right)\,;
  \label{eq:qtij}
\end{equation}
and the probability of node~$i$ not being infected by any 2-simplex contatining node~$j$,
\begin{equation}
  u_{ij}^\triangle(t) = \prod_{r \in \triangle_{ij}}
                        \left(1 - \beta^\triangle \frac{P_{ijr}^{SII}(t)}{P_{ij}^{SI}(t)}\right)\,.
  \label{eq:uij}
\end{equation}

\begin{figure}[t]
  \includegraphics[width=\columnwidth]{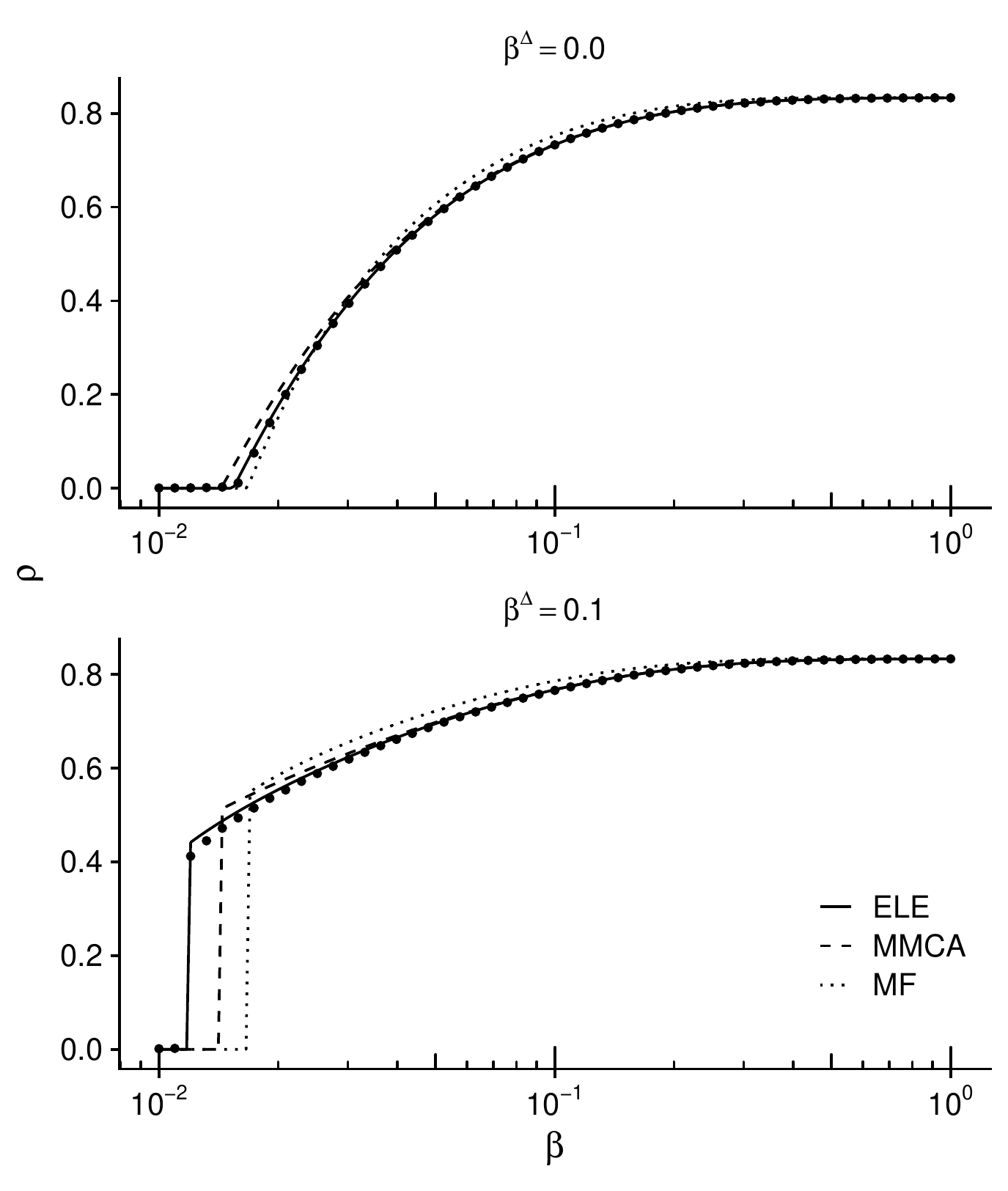}
  \caption{Incidence of the epidemic $\rho$ as function of infection probability $\beta$ on a random network of $N = 2000$ nodes. The connectivity of the network and the epidemic parameters are the same as those specified in Figure~\ref{fig:mf_stability}. The incidence has been analytically computed using ELE (solid line), MMCA (dashed line) and the mean-field approximation (dotted line). Results obtained from Monte Carlo simulations, performed using the quasistationary approach~\cite{ferreira}, are depicted by solid dots.}
  \label{fig:betarho_er_bw}
\end{figure}


\begin{figure}[t]
  \includegraphics[width=\columnwidth]{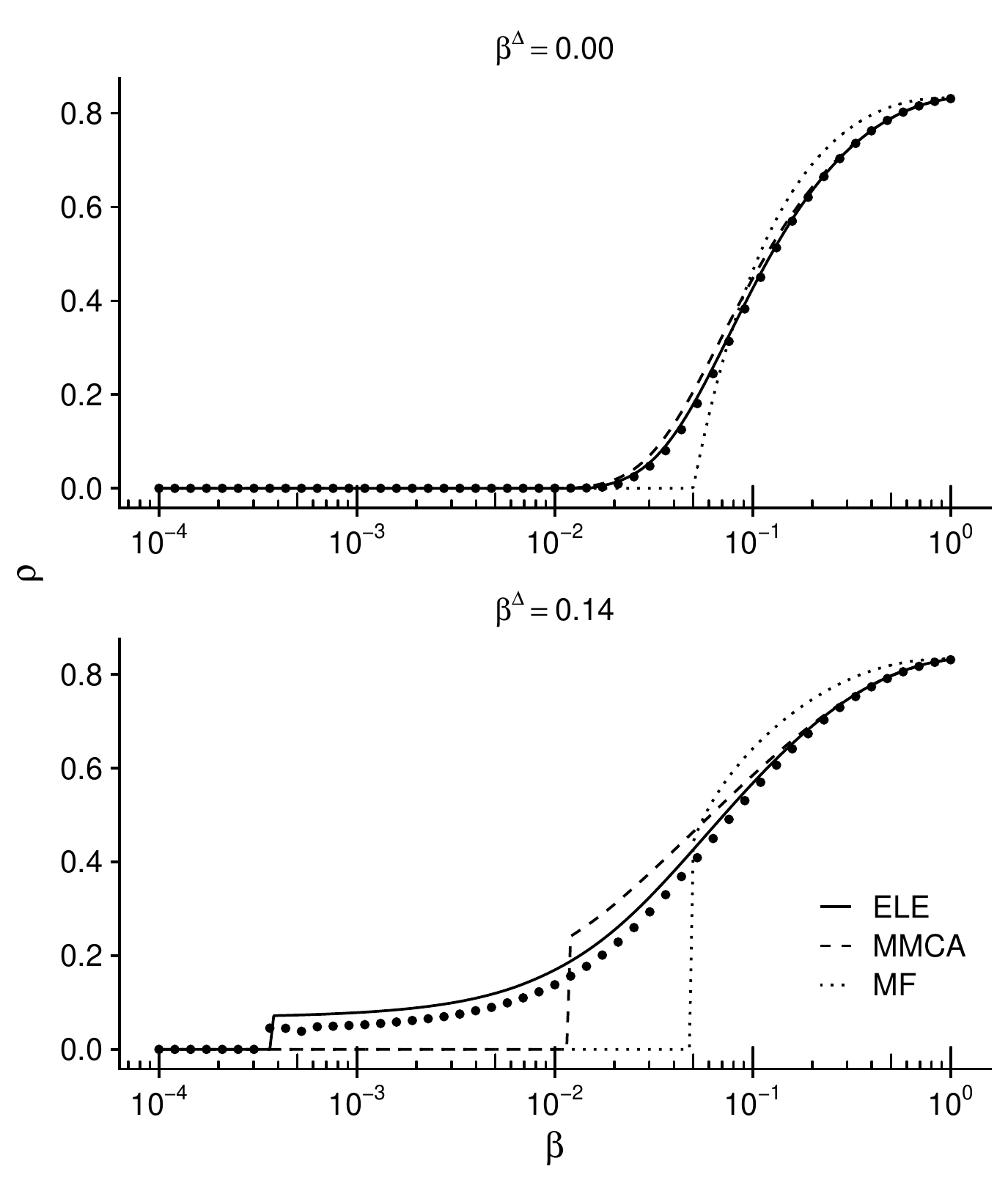}
  \caption{Incidence of the epidemic $\rho$ as function of infection probability $\beta$ on a random network with $N = 8000$ nodes, built according to~\cite{doro02}. The average degree of the network is $\langle k \rangle = 4$ and the average number of triangles per node is $\langle k^{\triangle} \rangle = 3$. The recovery probability is $\mu = 0.2$. The incidence has been analytically computed using ELE (solid line), MMCA (dashed line) and the mean-field approximation (dotted line). Results obtained from Monte Carlo simulations, performed using the quasistationary approach~\cite{ferreira}, are depicted by solid dots.}
  \label{fig:betarho_dorogo_v3_bw}
\end{figure}

See an illustration of the different contributions to Eqs.~(\ref{eq:qij}) to~(\ref{eq:uij}) in Figure~\ref{fig:schema_bw}. To solve the system of Eqs.~(\ref{eq:nodesELE}) and~(\ref{eq:linksELE}), we still need a closure for the ternary joint probabilities $P_{ijr}^{SII}$ found in Eqs.{(\ref{eq:qtij}) and~(\ref{eq:uij}). To break the hierarchy of clusters produced by these terms we must rely on approximations. The classical pair approximation in statistical physics~\cite{henkel}, also used in the context of epidemics by Mata and Ferrerira~\cite{mata}, would consist in approximating $P_{ijr}^{SII} \approx P_{ij}^{SI}P_{jr}^{II}/P_{j}^{I}$. However, Cator and Van Mieghem~\cite{cator}, propose a different closure $P_{ijr}^{SII} \approx P_{ij}^{SI}P_{r}^{I}$. The main problem with these proposals in the current scenario is that the symmetry of the elements in the 2-simplexes is broken, and this concurs in a degeneration of solutions. For example, using the second option, we could choose between three different closures for $P_{ijr}^{SII}$: $P_{ij}^{SI}P_{r}^{I}$, $P_{ir}^{SI}P_{j}^{I}$, or $P_{jr}^{II}P_{i}^{S}$. Each alternative makes use of one node and one link (the link opposite to the chosen node in the triangle) probabilities, but there is no indication in the structure of which ones should be preferred. To avoid such degeneracies, the closure proposal should be symmetric, and hence our approach consists in the following closure approximation:
\begin{equation}
  \label{eq:closure}
  P_{ijr}^{SII} = \frac{P^{SI}_{ij}P^{SI}_{ir}P^{II}_{jr}}{P^S_i P^I_j P^I_r}\,.
\end{equation}
In this way, all the node and link probabilities of the 2-simplex structures are used, avoiding the asymmetries introduced in the two previous closures, which were designed to handle connected triads of nodes, not necessarily forming triangles as in our current case.

The results of the previous mathematical formulation can now be obtained by fixed point iteration. In Figure~\ref{fig:betarho_er_bw} we present the results for homogeneous random networks, when the simplicial structure is not considered ($\beta^{\triangle}=0$), and when 2-simplices are included ($\beta^{\triangle}\neq0$). We observe that in the second case, when the simplical structure is considered for the higher-order dynamics of the SIS model, the incidence of the epidemics $\rho=\frac{1}{N}\sum_i P_i^{I}$ reveals an abrupt transition in all the previous approaches: MMCA, its mean-field version MF, and ELE. Nevertheless, the most accurate approximation when compared with the Monte Carlo simulation of the system, is provided by the epidemic link equations ELE. This is specially crucial when we try to capture the critical point of the transition. Unfortunately, the critical point at which the transition occurs eludes our analytical determination in ELE, given that nor the usual linearization technique \cite{gomez10,matamalas18}, neither the next generation matrix method (NGM)~\cite{dickman} provided meaningful results. In Figure~\ref{fig:betarho_dorogo_v3_bw} we corroborate our results on heterogeneous (scale-free) networks constructed following the analytical proposal in~\cite{doro02}. We observe how heterogeneity in degree highlights the differences in the determination of the critical point obtained by ELE, MMCA, and MF.

Summarizing, we have presented the mathematical formulation of the SIS model in simplicial complexes, using a discrete time probabilistic description of the process, in the node approximation MMCA, and in the link approximation ELE. Both descriptions predict an abrupt transition, as well as the stationary homogeneous approximation of the MMCA, the MF. The accuracy of the predictions is largely better for ELE, and reveals that this approximation is extremely useful when dealing with the simplicial geometry of complex networks. For the determination of the critical points of ELE, we think that further analysis using ideas on stability of subsystems~\cite{perc} of 1--simplices and 2--simplices, are a promising line of research. The results are not only important for epidemic spreading, but for any other contagion process that can be described within the probabilistic framework of MMCA.

We acknowledge support by Ministerio de Econom\'{i}a y Competitividad (Grants No.\ PGC2018-094754-B-C21 and No.\ FIS2015-71929-REDT), Generalitat de Catalunya (Grant No.\ 2017SGR-896), and Universitat Rovira i Virgili (Grant No.\ 2018PFR-URV-B2-41). AA acknowledges also ICREA Academia and the James S. McDonnell Foundation (Grant No.\ 220020325).



\end{document}